\documentclass{easychair}

\usepackage{microtype}

\begin{document}

%
\title{CPC: programming with a massive number of lightweight threads}

%
\titlerunning{CPC: programming with a massive number of lightweight threads}


%
\author{Gabriel Kerneis\\
Universit\'e Paris Diderot\\
Paris, France\\
\url{kerneis@pps.jussieu.fr}\\
\and
Juliusz Chroboczek\\
Universit\'e Paris Diderot\\
Paris, France
}

%
\authorrunning{Kerneis, Chroboczek}

\maketitle

%

\section{Introduction}

Threads are a convenient and modular abstraction for writing concurrent
programs.  Unfortunately, threads, as they are usually implemented, are
fairly expensive, which often forces the programmer to use a somewhat
coarser concurrency structure than he would want to.  The standard
alternative to threads, event-loop programming, allows much lighter units
of concurrency; however, event-loop programming splits the flow of control
of a program into small pieces, which leads to code that is difficult to
write and even harder to understand \cite{adya,behren}.

\emph{Continuation Passing C} (CPC) \cite{chroboczek,cpc2010} is a
translator that converts a program written in threaded style into a
program written with events and native system threads, at the
programmer's choice.  Threads in CPC, when compiled to events, are
extremely cheap, roughly two orders of magnitude cheaper than in
traditional programming systems; this encourages a somewhat unusual
programming style.

Together with two undergraduate students \cite{attar-canal}, we taught
ourselves how to program in CPC by writing \emph{Hekate},
a \emph{BitTorrent} \emph{seeder}, a massively concurrent network server
designed to efficiently handle tens of thousands of simultaneously
connected peers.  In this paper, we describe a number of programming
idioms that we learnt while writing Hekate; while some of these idioms
are specific to CPC, many should be applicable to other programming
systems with sufficiently cheap threads.

The CPC translation process itself is described in detail elsewhere
\cite{cpc2010}.

\section{Cooperative CPC threads}
\label{sec:cpc-threads}

The extremely lightweight, cooperative threads of CPC lead to a ``threads
are everywhere'' feeling that encourages a somewhat unusual programming
style.

\paragraph{Lightweight threads}
Contrary to the common model of using one thread per client, Hekate
spawns at least three threads for every connecting peer: a reader, a
writer, and a timeout thread.  Spawning several CPC threads per client
is not an issue, especially when only a few of them are active at any
time, because idle CPC threads carry virtually no overhead.

The first thread reads incoming requests and manages the state of the
client.  The BitTorrent protocol defines two states for interested peers:
``unchoked,'' i.e.\ currently served, and ``choked.''  Hekate maintains
90\,\% of its peers in choked state, and unchokes them in a round-robin
fashion.

The second thread is in charge of actually sending the chunks of data
requested by the peer.  It usually sleeps on a condition variable,
and is woken up by the first thread when needed.  Because these
threads are scheduled cooperatively, the list of pending chunks is
manipulated by the two threads without need for a lock.

Each read on a network interface is guarded by a timeout, and a peer that
has not been involved in any activity for a period of time is disconnected.
Earlier versions of Hekate which did not include this protection would end
up clogged by idle peers, which prevented new peers from connecting.

In order to simplify the protocol-related code, timeouts are implemented in
the buffered read function, which spawns a new timeout thread on each
invocation.  This temporary third thread sleeps for the duration of the timeout,
and aborts the I/O if it is still pending.  Because most timeouts do not
expire, this solution relies on the efficiency of spawning and
context-switching short-lived CPC threads \cite{chroboczek,cpc2010}.

\paragraph{Native and cps functions}
CPC threads might execute two kinds of code: \emph{native} functions
and \emph{cps} functions (annotated with the \texttt{cps} keyword).
Intuitively, cps functions are interruptible and native functions are not.
From a more technical point of view, cps functions are compiled by
performing a transformation to Continuation Passing Style (CPS), while
native functions execute on the native stack \cite{cpc2010}.

There is a global constraint on the call graph of a CPC program: a cps
function may only be called by a cps function; equivalently, a native
function can only call native functions --- but a cps function can call
a native function.  This means that at any point in time, the dynamic
chain consists of a ``cps stack'' of cooperating functions followed by a
``native stack'' of regular C functions.  Since context switches are
forbidden in native functions, only the former needs to be saved and
restored when a thread cooperates.

Figure~\ref{fig:listening} shows an example of a cps function:
\texttt{listening} calls the primitive \texttt{cpc\_io\_wait} to wait
for the file descriptor \texttt{socket\_fd} to be ready, before
accepting incoming connections with the native function \texttt{accept}
and spawning a new thread for each of them.

\begin{figure}
\begin{center}
\begin{minipage}{0.5\linewidth}
\small
\begin{verbatim}
cps void
listening(hashtable * table) {
    /* ... */
    while(1) {
        cpc_io_wait(socket_fd, CPC_IO_IN);
        client_fd = accept(socket_fd, ...);
        cpc_spawn client(table, client_fd);
    }
}
\end{verbatim}
\end{minipage}
\end{center}
\caption{Accepting connections and spawning threads}
\label{fig:listening}
\end{figure}

\section{Comparison with event-driven programming}
\label{sec:comparison}

\paragraph{Code readability}

Hekate's code is much more readable than its event-driven equivalents.
Consider for instance the BitTorrent handshake, a message exchange
occurring just after a connection is established.  In
\emph{Transmission}\footnote{\url{http://www.transmissionbt.com}},
a popular and efficient BitTorrent client written in (mostly)
event-driven style, the handshake is a complex piece of code, spanning
over a thousand lines in a dedicated file.  By contrast, Hekate's
handshake is a single function of less than fifty lines including error
handling.

While some of Transmission's complexity is explained by its support for
encrypted connexions, Transmission's code is intrinsically much more
messy due to the use of callbacks and a state machine to keep track of
the progress of the handshake.  This results in an obfuscated flow of
control, scattered through a dozen of functions (excluding
encryption-related functions), typical of event-driven code \cite{adya}.

\paragraph{Expressivity}

Surprisingly enough, CPC threads turn out to be more expressive than native
threads, and allow some idioms that are more typical of event-driven style.

A case in point: buffer allocation for reading data from the network.
When a native thread performs a blocking read, it needs to allocate the
buffer before the \texttt{read} system call; when many threads are
blocked waiting for a read, these buffers add up to a significant amount
of storage.  In an event-driven program, it is possible to delay
allocating the buffer until after an event indicating that data is
available has been received.

The same technique is not only possible, but actually natural in CPC:
buffers in Hekate are only allocated after \texttt{cpc\_io\_wait} has
successfully returned.  This provides the reduced storage requirements
of an event-driven program while retaining the linear flow of control of
threads.

\section{Detached threads}
\label{sec:detached-threads}

While cooperative, deterministically scheduled threads are less error-prone
and easier to reason about than preemptive threads, there are circumstances
in which native operating system threads are necessary.  In traditional
systems, this implies either converting the whole program to use native
threads, or manually managing both kinds of threads.

A CPC thread can switch from cooperative to preemptive mode at
any time by using the the \texttt{cpc\_attach} primitive (inspired by
FairThreads' \texttt{ft\_thread\_link} \cite{boussinot}).  A cooperative
thread is said to be \emph{attached} to the default scheduler, while
a preemptive one is \emph{detached}.

The \texttt{cpc\_attach} primitive takes a single argument, a scheduler,
either the default event loop (for cooperative scheduling) or a thread pool
(for preemptive scheduling).  It returns the previous scheduler, which
makes it possible to eventually restore the thread to its original state.
Syntactic sugar is provided to execute a block of code in attached or
detached mode (\texttt{cpc\_attached}, \texttt{cpc\_detached}).

Hekate is written in mostly non-blocking cooperative style; hence, Hekate's
threads remain attached most of the time.  There are a few situations,
however, where the ability to detach a thread is needed.

\paragraph{Blocking OS interfaces}

Some operating system interfaces, like the \texttt{getaddrinfo} DNS
resolver interface, may block for a long time (up to several seconds).
Although there exist several libraries which implement equivalent
functionality in a non-blocking manner, in CPC we simply enclose the call
to the blocking interface in a \texttt{cpc\_detached} block (see
Figure~\ref{fig:cpc-detached}a).

Figure~\ref{fig:cpc-detached}b shows how \texttt{cpc\_detached} is
expanded by the compiler into two calls to \texttt{cpc\_attach}.  Note that
CPC takes care to attach the thread before returning to the caller
function, even though the \texttt{return} statement is inside the
\texttt{cpc\_detached} block.

\begin{figure}[htb]
\small\centering
\begin{tabular}{ l | l }
&
\verb+cpc_scheduler *s =+\\
\verb+cpc_detached {+&
\verb+    cpc_attach(cpc_default_threadpool);+\\
\verb+    rc = getaddrinfo(name, ...)+&
\verb+rc = getaddrinfo(name, ...)+\\
\verb+    return rc;+&
\verb+cpc_attach(s);+\\
\verb+}+&
\verb+return rc;+\\
\multicolumn{1}{c}{\normalsize(a)}&
\multicolumn{1}{c}{\normalsize(b)}
\end{tabular}
\normalsize
\caption{Expansion of \texttt{cpc\_detached} in terms of \texttt{cpc\_attach}}
\label{fig:cpc-detached}
\end{figure}

\paragraph{Blocking library interfaces}
Hekate uses the \textit{curl} library
\footnote{\url{http://curl.haxx.se/libcurl/}} to contact BitTorrent
\emph{trackers} over HTTP.  Curl offers both a simple, blocking
interface and a complex, non-blocking one.  We decided to use the one
interface that we actually understand, and therefore call the blocking
interface from a detached thread.

\paragraph{Parallelism}
Detached threads make it possible to run on multiple processors or
processor cores.  Hekate does not use this feature, but a CPU-bound program
would detach computationally intensive tasks and let the kernel schedule
them on several processing units.

\section{Hybrid programming}
\label{sec:hybrid-threads}

Most realistic event-driven programs are actually \emph{hybrid} programs
\cite{pai,welsh}: they consist of a large event loop, and a number of
threads (this is the case, by the way, of the \emph{Transmission}
BitTorrent client mentioned above).  Such blending of native threads
with event-driven code is made very easy by CPC, where switching from
one style to the other is a simple matter of using the \verb|cpc_attach|
primitive.

This ability is used in Hekate for dealing with disk reads.  Reading from
disk might block if the data is not in cache; however, if the data is
already in cache, it would be wasteful to pay the cost of a detached
thread.  This is a significant concern for a BitTorrent seeder because
the protocol allows requesting chunks in random order, making kernel
readahead heuristics useless.

\begin{figure}
\begin{center}
\begin{minipage}{0.8\linewidth}
\small
\begin{verbatim}
    prefetch(source, length);                           /* (1) */
    cpc_yield();                                        /* (2) */
    if(!incore(source, length)) {                       /* (3) */
        cpc_yield();                                    /* (4) */
        if(!incore(source, length)) {                   /* (5) */
            cpc_detached {                              /* (6) */
                rc = cpc_write(fd, source, length);
            }
            goto done;
        }
    }
    rc = cpc_write(fd, source, length);                 /* (7) */
done:
    ...
\end{verbatim}
\end{minipage}
\end{center}
{\small\em The functions \verb|prefetch| and \verb|incore| are thin
wrappers around the {\tt posix\_madvise} and {\tt mincore} system
calls.}
\caption{An example of hybrid programming (non-blocking read)}
\label{fig:non-blocking-read}
\end{figure}

The actual code is shown in Figure~\ref{fig:non-blocking-read}: it sends
a chunk of data from a memory-mapped disk file over a network socket.
In this code, we first trigger an asynchronous read of the on-disk data
(1), and immediately yield to threads servicing other clients (2) in
order to give the kernel a chance to perform the read.  When we are
scheduled again, we check whether the read has completed (3); if it has,
we perform a non-blocking write (7); if it hasn't, we yield one more
time (4) and, if that fails again (5), delegate the work to a native
thread which can block (6).

Note that this code contains a race condition: the prefetched block of
data could have been swapped out before the call to {\tt cpc\_write},
which would stall Hekate until the write completes.  However, our
measurements show that the write never lasted more than 10~ms, which
clearly indicates that the race does not happen.  Note further
that the call to {\tt cpc\_write} in the {\tt cpc\_detached} block (6)
could be replaced by a call to {\tt write}: we are in a native thread
here, so the non-blocking wrapper is not needed.  However, the CPC
runtime is smart enough to detect this case, and {\tt cpc\_write} simply
behaves as {\tt write} when invoked in detached mode; for simplicity, we
choose to use the CPC wrappers throughout our code.

\section{Experimental results}
\label{sec:experimental-results}

Benchmarking a BitTorrent seeder is a difficult task because it relies
either on a real-world load, which is hard to control and only provides
seeder-side information, or on an artificial testbed, which might fail
to accurately reproduce real-world behaviour.  Our experience with
Hekate in both kinds of setup shows that CPC generates efficient code,
lightweight enough to run Hekate on embedded hardware.  This confirms
our earlier results \cite{kerneis}, where me measured the performance of
toy web servers.

\paragraph{Real-world workload}
To benchmark the ability of Hekate to sustain a real-world load, we need
popular torrents with many requesting peers over a long period of time.
Updates for Blizzard's game \textit{World of Warcraft} (WoW), distributed
over BitTorrent, meet those conditions: each of the millions of WoW players
around the world runs a hidden BitTorrent client, and at any time many of
them are looking for the latest update.

We have run an instance of Hekate seeding WoW updates without interruption
for weeks.  We saw up to 1,000 connected peers (800 on average) and
a throughput of up to 10\,MB/s (around 5\,MB/s on average).
Hekate never used more than 10\,\% of the 3.16\,GHz dual core CPU of our
benchmarking machine.

\paragraph{Stress-test on embedded hardware}
We have ported Hekate to \emph{OpenWrt}\footnote{\url{http://openwrt.org}},
a Linux distribution for embedded devices.  Hekate runs flawlessly on
a MIPS-based router with a 266\,MHz CPU, 32\,MB of RAM and a 100\,Mbps
network card.  The torrent files were kept on a USB key.

Because Hekate maps every file it serves in memory, and the MIPS routers
running OpenWrt are 32-bit machines, we are restricted to no more than 2\,GB
of content.  Our stress-test consists in 1,000 clients, requesting random
chunks of a 1.2\,GB torrent from a computer directly connected to the
device.  Hekate sustained a throughput of 2.9\,MB/s.  The CPU was
saturated, mostly with software interrupt requests (60\,\% \textit{sirq},
the usb-storage kernel module using up to 25\,\% of CPU).

\section{Conclusions}

Hekate has shown that CPC is a tool that is able to produce efficient
network servers, even when used by people who do not fully understand
its internals and are not specialists of network programming.  While
writing Hekate, we had a lot of fun exploring the somewhat unusual
programming style that CPC's lightweight, hybrid threads encourage.

We have no doubt that CPC, possibly with some improvements, will turn
out to be applicable to a wider range of applications than just network
servers, and are looking forward to experimenting with CPU-bound
distributed programs.

%
\label{sect:bib}

\end{document}